\documentclass[aps,prd,twocolumn,showpacs,10pt,superscriptaddress,preprintnumbers,nofootinbib]{revtex4-1}
\usepackage{amsmath}
\usepackage{physics, bm}
\usepackage{graphicx}
\usepackage{amssymb}
\usepackage[colorlinks,citecolor=red]{hyperref}

\usepackage{color}    
\usepackage{ulem}
\usepackage{multirow}

\newcommand{\calg}{\mathcal{G}}

\newcommand{\calO}{ {\cal O} }
\newcommand{\re}{{\rm Re}}
\newcommand{\im}{{\rm Im}}

\begin{document}

\title{Probing the four-fermion operators via the transverse double spin asymmetry at the Electron-Ion Collider}

\author{Hao-Lin Wang}
\email{whaolin@m.scnu.edu.cn}
\affiliation{Key Laboratory of Atomic and Subatomic Structure and Quantum Control (MOE), Guangdong Basic Research Center of Excellence for Structure and Fundamental Interactions of Matter, Institute of Quantum Matter, South China Normal University, Guangzhou 510006, China}
\affiliation{Guangdong-Hong Kong Joint Laboratory of Quantum Matter, Guangdong Provincial Key Laboratory of Nuclear Science, Southern Nuclear Science Computing Center, South China Normal University, Guangzhou 510006, China}

\author{Xin-Kai Wen}
\email{xinkaiwen@pku.edu.cn}
\affiliation{School of Physics, Peking University, Beijing 100871, China}

\author{Hongxi Xing}
\email{hxing@m.scnu.edu.cn}
\affiliation{Key Laboratory of Atomic and Subatomic Structure and Quantum Control (MOE), Guangdong Basic Research Center of Excellence for Structure and Fundamental Interactions of Matter, Institute of Quantum Matter, South China Normal University, Guangzhou 510006, China}
\affiliation{Guangdong-Hong Kong Joint Laboratory of Quantum Matter, Guangdong Provincial Key Laboratory of Nuclear Science, Southern Nuclear Science Computing Center, South China Normal University, Guangzhou 510006, China}
\affiliation{Southern Center for Nuclear-Science Theory (SCNT),
Institute of Modern Physics, Chinese Academy of Sciences, Huizhou 516000, China}
 
\author{Bin Yan}
\email{yanbin@ihep.ac.cn (corresponding author)}
\affiliation{Institute of High Energy Physics, Chinese Academy of Sciences, Beijing 100049, China}

\begin{abstract}
\noindent
The chirality-flipping operators of light fermions are currently poorly constrained by experimental analyses due to the lack of interference with Standard Model (SM) amplitudes in traditional observables. In this work, we propose to investigate the semi-leptonic scalar/tensor four-fermion operators of electron and quarks through the transverse double spin asymmetry (DSA) at Electron-Ion Collider, where both the electron and proton beams could be highly transversely polarized. Due to the chirality-flipping nature of these operators, we demonstrate that their interference with the SM results in an unsuppressed contribution to the DSA, and could lead to non-trivial azimuthal $\cos2\phi$ and $\sin2\phi$ distributions that are linearly dependent on their Wilson coefficients.  This new method has the potential to significantly improve the current constraints on these scalar/tensor four-fermion operators without relying on theoretical assumptions about other types of new physics effects, particularly for the tensor type operator of the $u$-quark. Additionally, our findings indicate that both the real and imaginary parts of these operators can be simultaneously constrained and offer a new opportunity for probing potential $CP$-violation effects. However, it is important to note that these results would be sensitive to the quark transversity distributions, which are currently poorly constrained by the experimental data, but could be significantly improved at the upcoming Electron-Ion Collider. Therefore, our work opens up a new
avenue to utilize this new spin asymmetry for exploring the new physics effects from the scalar/tensor four-fermion operators.
\end{abstract}

\maketitle

\section{Introduction}
\label{sec:intro}
The absence of signals for new heavy resonances at the Large Hadron Collider (LHC) strongly suggests that the scale ($\Lambda$) of new physics (NP) is likely to be significantly larger than the electroweak scale. In light of this, the standard model effective field theory (SMEFT) has emerged as a powerful theoretical framework for systematically parametrizing potential NP effects. This is achieved by introducing a series of effective interactions involving higher-dimensional operators, which are constructed from the dynamical degrees of freedom of the standard model (SM) with the gauge symmetry $SU(3)_C\otimes SU(2)_L\otimes U(1)_Y$. It has been demonstrated that the operators with odd dimension can lead to violations of lepton number ($L$) and/or baryon number ($B$)~\cite{Kobach:2016ami}. Therefore, when assuming the conservation of $B$ and $L$, the leading contributions to the observables from NP are expected to originate from dimension-6 (dim-6) operators of the form $C_i \mathcal{O}_i/\Lambda^2$, which have been systematically constructed in Refs.~\cite{Buchmuller:1985jz,Grzadkowski:2010es}. The dimensionless Wilson coefficient $C_i$ describes the interaction strength of the operator $\calO_i$.

There have been intensive theoretical and experimental efforts in recent years to constrain those subsets of the dim-6 operators with the measurements of total cross sections and differential distributions of SM processes at the LHC and other facilities, see, e.g.~\cite{Englert:2014uua,Falkowski:2015fla,Corbett:2015ksa,Cao:2015doa,Cao:2015qta,Cao:2015oaa,Cao:2016zob,Cirigliano:2016nyn,Alioli:2018ljm,Durieux:2018tev,Degrande:2018fog,Vryonidou:2018eyv,Durieux:2018ggn,Cao:2018cms,DeBlas:2019qco,Brivio:2019ius,Hartland:2019bjb,Du:2020dwr,Alioli:2020kez,Cirigliano:2021img,Ethier:2021bye,Miralles:2021dyw,Yan:2021tmw,Boughezal:2021tih,Cirigliano:2021peb,Cao:2021wcc,Du:2021rdg,Liao:2021qfj,Liu:2022vgo,deBlas:2022ofj,Dawson:2022bxd,Greljo:2022cah,Grunwald:2023nli,Cao:2023qks,Wen:2023xxc,Shao:2023bga,Chai:2024zyl}. These studies have yielded valuable insights and have significantly limited the impact of many NP effects associated with these dim-6 operators. However, there are certain types of dim-6 operators that remain poorly constrained. These include the chirality-flipping operators, which consist of four-fermion operators with scalar or tensor structures, Yukawa-like operators, and dipole operators involving light fermions. The contributions of these operators to cross sections through the interference with the SM are suppressed by the negligible mass of the light fermion at $\calO(1/\Lambda^2)$. Consequently, their leading contributions to unpolarized observables, as reported in the literature, are typically at the order of $\calO(1/\Lambda^4)$~\cite{Escribano:1993xr,daSilvaAlmeida:2019cbr,Boughezal:2021tih,Cao:2021trr,Boughezal:2023ooo,Grunwald:2023nli}.
However, it is crucial to precisely measure these NP effects in order to gain a deeper understanding of NP beyond the SM. One notable example is the recent exciting news from the muon $g-2$ measurement at Fermilab, which has revealed a significant deviation between experimental data and SM predictions at a 5$\sigma$ significance level~\cite{Muong-2:2023cdq}. This intriguing discrepancy could potentially be explained by the presence of dipole operators arising from NP. Similarly,  the scalar and tensor four-fermion operators may be induced by an additional scalar or gauge boson, making them the key players in unraveling the nature of the underlying theory~\cite{Li:2023cwy}. Therefore, it is imperative to undertake precise measurements of the Wilson coefficients associated with these operators using current and future experimental facilities. 

Recently, it has been demonstrated that the electron dipole operators can be probed through the single transverse spin asymmetry (SSA) at $\calO(1/\Lambda^2)$, without the suppression of the electron mass~\cite{Wen:2023xxc}. This can be achieved by studying the interference of the electron dipole operators and the SM at a future lepton collider with transversely polarized lepton beams. The prospects for probing these operators at such a collider are promising, as the resulting limits on the Wilson coefficients can be improved by one to two orders of magnitude compared to current measurements at the LHC and LEP~\cite{Wen:2023xxc}. This idea can also be extended to the forthcoming Electron-Ion Collider (EIC) and the planned EIC in China (EicC), where high polarizations of electron and proton beams can be achieved~\cite{Boughezal:2023ooo}. These facilities were initially designed to precisely determine the spin-dependent parton distribution functions (PDFs) and explore the spin and 3D structure of the nucleon~\cite{AbdulKhalek:2021gbh,Anderle:2021wcy}. However, it has been demonstrated that these facilities also have the potential to probe the electroweak properties of the SM and search for potential NP effects~\cite{AbdulKhalek:2022hcn,Boughezal:2020uwq,Li:2021uww,Yan:2021htf,Boughezal:2023ooo,Yan:2022npz,Davoudiasl:2021mjy,Liu:2021lan,Batell:2022ogj,Davoudiasl:2023pkq,Cirigliano:2021img,Zhang:2022zuz,Gonderinger:2010yn,Boughezal:2022pmb,Balkin:2023gya}. The high polarization of the beams at the EIC/EicC opens up new avenues for studying these operators and improving our understanding of the underlying physics beyond the SM. 

In this paper, we extend the analysis of electron and quark dipole operators in Refs.~\cite{Wen:2023xxc,Boughezal:2023ooo} and explore the possibilities of probing the scalar and tensor type four-fermion operators by considering the transverse double spin asymmetry (DSA) of the electron and proton beams in inclusive deeply-inelastic scattering (DIS) at the EIC/EicC. We demonstrate that, similar to the dipole operators, the interference between the scalar/tensor type four-fermion operators and the SM results in nontrivial azimuthal $\cos2\phi$ and $\sin2\phi$ distributions for $\calO_{l e d q}$ and only flat distributions for $\calO_{l e q u}^{(1,3)}$ with aligned/opposite spin setups of the electron and proton. These distributions are switched between the real and imaginary couplings if the spin configurations become perpendicular. Importantly, all of these distributions are linearly dependent on the Wilson coefficients associated with these operators at $\calO(1/\Lambda^2)$, without any suppression from the electron and quark masses. Furthermore, the results are not significantly affected by the presence of other NP operators in the DIS process, such as the vector and axial-vector type four-fermion operators. 

The paper is organized as follows. In Section \ref{sec:DSA-SM}, we calculate the DSA for the DIS process in the SM, and provide the numerical estimation at the EIC. We then calculate the DSA modified by the four-fermion operators in the SMEFT in Section \ref{sec:DSA-SMEFT}. We show the enhancement and azimuthal distribution for the DSA with the chirality-flipping operators. The expected sensitivities of probing the four-fermion operators through the transverse DSA at EIC and EicC are given in Section \ref{sec:sensitivity}. Additionally, we discuss the effects of the quark transversity distributions and the constraints for such operators from other processes. Our concluding remarks are presented in Section \ref{sec:conclusion}.

\section{Transverse double spin asymmetry in the SM}
\label{sec:DSA-SM}
In this section, we calculate the transverse DSA for the SM inclusive DIS process $e^-(k)+p(P) \rightarrow e^-(k^\prime)+X$ at the EIC/EicC. The transverse spin vectors of the electron and proton can be expressed as,
\begin{align}
	\nonumber
	S_{T,e}^\mu=P_{T,e}(0,\cos\phi_1,\sin\phi_1,0),
	\\
	S_{T,p}^\mu=P_{T,p}(0,\cos\phi_2,\sin\phi_2,0),
\end{align}
where $P_{T,e}~(P_{T,p})$ represents the magnitude of the electron's (proton's) transverse polarization, and $\phi_1~(\phi_2)$ is the angle between the transverse spin of the incoming lepton (proton) and the momentum of the outgoing electron in the transverse plane. The transverse DSA is defined as:
\begin{align}
	A_{TT}=\frac{\sigma\left(e^{\uparrow} p^{\uparrow}\right)+\sigma\left(e^{\downarrow} p^{\downarrow}\right)-\sigma\left(e^{\uparrow} p^{\downarrow}\right)-\sigma\left(e^{\downarrow} p^{\uparrow}\right)}{\sigma\left(e^{\uparrow} p^{\uparrow}\right)+\sigma\left(e^{\downarrow} p^{\downarrow}\right)+\sigma\left(e^{\uparrow} p^{\downarrow}\right)+\sigma\left(e^{\downarrow} p^{\uparrow}\right)},
\end{align}
where the superscripts $\uparrow$ and $\downarrow$ indicate the directions of transverse spin of electron and proton with $P_{T,e}=P_{T,p}=1$.

In the SM, the transverse DSA can arise from the single-photon exchange process, given by
\begin{align}
	\label{eq:attsm:lo}
	\nonumber
	A_{TT}^{\text{SM},\gamma}
    &=\frac{2y^2 \left[(1-y)\cos\phi_+-(1+y)\cos\phi_-\right]}{Q^2} 
    \\
	&\times\frac{  \sum_q m_e m_q Q_q^2 h_q(x,\mu) }{\sum_q  f_q(x,\mu)\Big[Q_q^2(y^2-2y+2)-\mathcal{F}^{eq}_{Z}(Q^2)\Big]},
\end{align}
where $\phi_+\equiv\phi_1+\phi_2$, $\phi_-\equiv\phi_1-\phi_2$, and $f_q(x,\mu)$ and $h_q(x,\mu)$ respectively denote the PDF and transversity distribution of a quark with flavor $q$ and factorization scale $\mu=Q$ under leading-twist collinear factorization, and $Q_q$ is the corresponding electric charge of quark $q$. The kinematic variables in Eq.~\eqref{eq:attsm:lo} are defined as
\begin{align}
	Q^2= -\widetilde{q}^2=xy S,\quad
	x= \frac{Q^2}{2 P\cdot \widetilde{q}},\quad
	y= \frac{P\cdot \widetilde{q}}{P\cdot k},
\end{align}
where $\widetilde{q}=k-k^\prime$ denotes the momentum transfer of the electrons and $S=(k+P)^2$ is the center of mass energy square. The additional correction from $\gamma$-$Z$ interference in the unpolarized cross section has been encoded by the factor $\mathcal{F}^{eq}_{Z}$, which can not be ignored for the kinematic region of EIC,
\begin{align}
    \mathcal{F}^{eq}_{Z}(Q^2)\equiv 2\frac{\Tilde{\epsilon}_Q }{s_W^2 c_W^2}Q_q\Big[(y^2-2y+1)\calg_{-}^{eq}+\calg_{+}^{eq}\Big],
\end{align}
where $\Tilde{\epsilon}_Q$ is defined as
\begin{align}
    \Tilde{\epsilon}_Q\equiv\frac{Q^2}{Q^2+m_Z^2}.
\end{align}
Here, we have defined
\begin{align}
    \mathcal{G}_+^{eq} \equiv g_V^e g_V^q+g_A^e g_A^q,\quad
    \mathcal{G}_-^{eq} \equiv g_V^e g_V^q-g_A^e g_A^q
\end{align}
for abbreviation. The vector and axial vector couplings of the $Z$-boson to the fermion $f$ in the SM are given by
\begin{align}
	g_V^f=\frac{T_3^f}{2}-Q_f s_W^2 ,\quad g_A^f=-\frac{T_3^f}{2},
\end{align}
where $s_W\equiv \sin \theta_W$ with $\theta_W$ being the weak mixing angle, and 
\begin{align}
\left(T_3^f,~Q_f\right)=\left(\frac{1}{2},~\frac{2}{3}\right), \left(-\frac{1}{2},~ -\frac{1}{3}\right), \left(-\frac{1}{2},~ -1\right) 
\end{align}
for up-type, down-type quarks and electron, respectively. 
It is important to note that $m_q\sim 300~{\rm MeV}$ represents the typical constituent quark mass, which arises from the non-perturbative effect of the spontaneous breaking of chiral symmetry~\cite{Afanasev:2007ii}. Consequently, the DSA in the SM is significantly suppressed by both the electron and quark masses. Additional subleading contribution to the DSA arises from the interference between $Z$ boson and the photon, and its effect should be comparable to the contribution from $\mathcal{F}^{eq}_{Z}$ in Eq.~\eqref{eq:attsm:lo}, given by
\begin{align}
	\label{eq:attsm:nlo}
	\nonumber
	&A_{TT}^{\text{SM},\gamma Z}
    =\frac{-4 y}{s_W^2 c_W^2(Q^2+m_Z^2)(y^2-2y+2)}\frac{1}{\sum_q Q_q^2 f_q(x,\mu)}
    \\
    \nonumber
    &\quad\times\sum_q m_e m_q Q_q h_q(x,\mu)\left[\cos\phi_+y(1-y)\calg_{-}^{eq}\right.
    \\
    &\qquad\qquad\left.-\cos\phi_-(1+y)\left((y-1)\calg_{-}^{eq}+\calg_{+}^{eq}\right)\right].
\end{align}
Note that we have ignored the correction from $\mathcal{F}^{eq}_{Z}$ in unpolarized cross section in Eq.~\eqref{eq:attsm:nlo} since its contribution is negligible. The results for the anti-quark processes can be obtained through Eqs.\,\eqref{eq:attsm:lo} and \eqref{eq:attsm:nlo}, by taking $g^{q}_A\rightarrow -g^{q}_A$ due to the charge-conjugated transformation of fermion bilinear currents, or equivalently, $\calg_{+}^{eq}\leftrightarrow \calg_{-}^{eq} $.

To estimate the effects of the transverse DSA in the SM, we utilize the quark transversity distributions from Refs.~\cite{Kang:2015msa,Zeng:2023nnb,Gamberg:2022kdb}. 
These distributions are obtained from global analyses of the single spin asymmetries (Collins and/or Sivers) in semi-inclusive hadron production in DIS (SIDIS) and di-hadron productions in semi-inclusive $e^-e^+$ annihilation processes. 
In these analyses, the sea quark transversity distributions were usually assumed to be zero due to the fact that quark transversity distributions do not mix with gluons in the evolution~\cite{Kang:2015msa}. Notably, the potential impact of the transversity distributions from $\bar u$ and $\bar d$ has been recently addressed in Refs.~\cite{Zeng:2023nnb,Gamberg:2022kdb}. 
The JAM Collaboration has incorporated data on Sivers asymmetry in Drell-Yan and pion production in proton-proton collisions.
In addition, they have utilized lattice QCD data on nucleon tensor charges to minimize the uncertainty in extracting transversities~\cite{Gamberg:2022kdb,Cocuzza:2023oam,Cocuzza:2023vqs}. We found that their latest extraction utilizing data on single transverse-spin asymmetries involving dihadron fragmentation~\cite{Cocuzza:2023oam,Cocuzza:2023vqs} yields DSA results similar to those from~\cite{Gamberg:2022kdb}, indicating a universal behavior of transversity across different types of observables and lattice QCD. 
To ensure consistency with the global analysis of the transversities, we use the {\tt CT14LO} unpolarized PDFs~\cite{Dulat:2015mca} for calculating $A_{TT}^{\rm SM}$ in accordance with the transversities from Refs.~\cite{Kang:2015msa,Zeng:2023nnb}. Meanwhile, we utilize {\tt JAM22-PDF-proton-nlo}~\cite{Cocuzza:2022jye} when using the transversity from the JAM Collaboration\cite{Gamberg:2022kdb}.

\begin{figure}[t!]
	\hspace{-0.5cm}
	\includegraphics[width=0.5\textwidth]{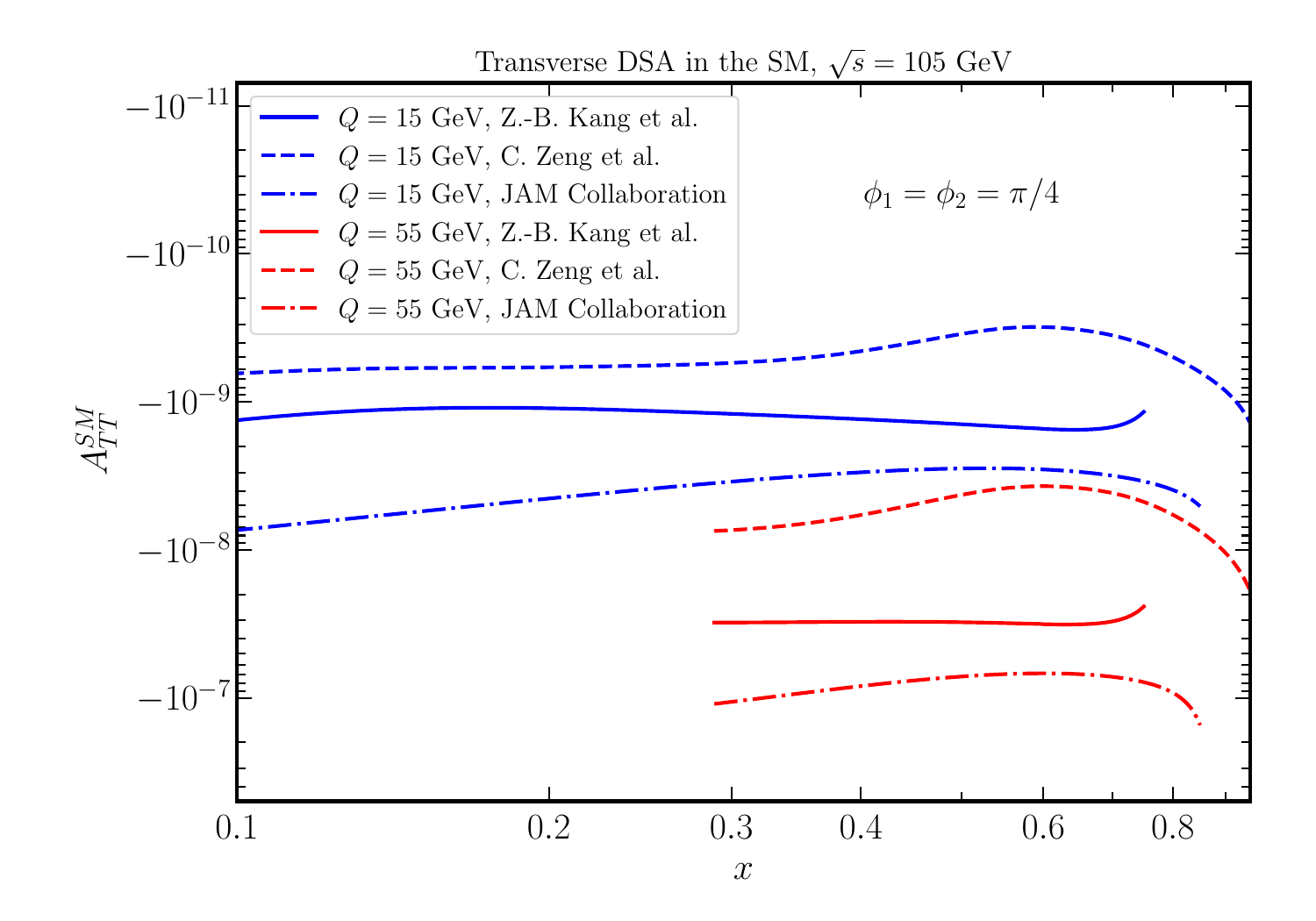}
	\caption{The transverse DSA $A_{TT}^{SM}$ in the SM from the leading-twist with $\phi_1=\phi_2=\pi/4$ as a function of $x$. The solid lines depict the results obtained using the transversities from Ref.\,\cite{Kang:2015msa}. The dashed lines correspond to the transversities from Ref.\,\cite{Zeng:2023nnb}, while the dot-dashed lines represent the transersities from Ref.\,\cite{Gamberg:2022kdb}.}
	\label{fig:SM}
\end{figure}

Figure~\ref{fig:SM} shows the transverse DSA in the SM ($A_{TT}^{\rm SM} = A_{TT}^{\rm SM,\gamma}+A_{TT}^{{\rm SM},\gamma Z}$) from the leading-twist with $\phi_1=\phi_2=\pi/4$ as a function of $x$ by assuming the momentum transfer $Q=15,~55~{\rm GeV}$ and $\sqrt{S}=105~{\rm GeV}$ at the EIC. Owing to the suppression
of the electron and quark masses, the $A_{TT}^{\rm SM}$ is estimated to be on the order of $ -(10^{-8}\sim10^{-9})$. 
The results are sensitive to the quark transversity distributions, which are poorly constrained by the current experimental data. It is important to note that the 
contributions associated with higher-twist PDFs cannot be ignored in this case compared to the predictions of the leading-twist. 
However, these results are suppressed by both the electron mass and $1/Q$, leading to the conclusion that the asymmetry in the SM remains negligible.
\footnote{The higher-order QCD corrections are expected to be negligible compared to the leading-order approximation in the SM, as shown in Ref.~\cite{deFlorian:2017ogw}}

\section{Transverse double spin asymmetry in the SMEFT}
\label{sec:DSA-SMEFT}
Now we investigate the transverse DSA induced by the scalar and tensor-type four-fermion operators from the SMEFT~\cite{Grzadkowski:2010es}~\footnote{The contributions from other NP operators will be suppressed by the mass of the electron and/or quark and can be ignored.},
\begin{align}
    \label{eq:ops}
		\nonumber
		\calO_{l e d q} &= \left(\bar{L}^j e\right)\left(\bar{d} Q^j\right), 
		\\ 
		\nonumber
		\calO_{l e q u}^{(1)} &= \left(\bar{L}^j e\right) \epsilon_{j k}\left(\bar{Q}^k u\right),
		\\
		\calO_{l e q u}^{(3)} &= \left(\bar{L}^j \sigma^{\mu \nu} e\right) \epsilon_{j k}\left(\bar{Q}^k \sigma_{\mu \nu} u\right),
\end{align}
where $L^j$ and $Q^j$ denote the $SU(2)_L$ doublets of lepton and quark fields with $j$ representing the $SU(2)_L$ index. The fields $e,u$ and $d$ are the $SU(2)_L$ singlets of charged lepton, up- and down-type quarks. The family indices of the fermion fields are suppressed. We should note that each operator $\calO_i$ is accompanied by a corresponding coefficient $C_i/\Lambda^2$, which parameterizes the short-distance ultraviolet (UV) physics.

Since these operators simultaneously flip the helicities of the electron and quark in the DIS process, the contributions from the operators in Eq.\,\eqref{eq:ops} to the transverse DSA will not be suppressed by the tiny quark and electron masses. The leading contributions to the DSA from them arise from the interference between the single-photon exchange SM amplitude and the NP, 
\begin{align}
	\label{DSA:SMEFT1}
	\nonumber
	\Delta A_{TT}^{\gamma}
    &=\frac{Q^2/4\pi\alpha}{\sum_q  f_q(x,\mu)\Big[Q_q^2(y^2-2y+2)-\mathcal{F}^{eq}_{Z}(Q^2)\Big]}
    \\
	\nonumber
	&\times\frac{1}{\Lambda^2}\Bigg(\sum_d Q_d h_d(x,\mu)(y-y^2) {\rm Re}\big[C_{ledq}e^{-i\phi_+}\big]
	\\
	\nonumber
	&~\qquad+\sum_u Q_u h_u(x,\mu)y {\rm Re}\big[C_{lequ}^{(1)}e^{-i\phi_-}\big] 
	\\
	&~\qquad+\sum_u Q_u h_u(x,\mu)4(y-2){\rm Re}\big[C_{lequ}^{(3)}e^{-i\phi_-}\big]\Bigg),
\end{align}
where $\alpha$ is the fine-structure constant. Again, we have included the correction $\mathcal{F}^{eq}_{Z}$ in Eq.~\eqref{DSA:SMEFT1} which arises from the $\gamma$-$Z$ interference in the unpolarized cross section. Additionally, the interference between $Z$-boson-exchange SM amplitude and the NP should be considered since its contribution is comparable to the effects of $\mathcal{F}^{eq}_{Z}$,
\begin{align}
	\label{DSA:SMEFT2}
	\nonumber
	\Delta A_{TT}^{Z}
   &=-\frac{1}{4\pi\alpha}\frac{Q^2}{y^2-2y+2}\frac{\Tilde{\epsilon}_Q}{s_W^2 c_W^2} \frac{1}{\sum_q Q_q^2 f_q(x,\mu)}
	\\
	\nonumber
	&\times \frac{1}{\Lambda^2}\Bigg(\sum_d  (y-y^2)\calg_-^{ed}h_d(x,\mu) {\rm Re}\big[C_{ledq}e^{-i\phi_+}\big]
	\\
	\nonumber
	&~\qquad+\sum_u y \calg_+^{eu} h_u(x,\mu) {\rm Re}\big[C_{lequ}^{(1)}e^{-i\phi_-}\big]
	\\
	&~\qquad+\sum_u 4(y-2)\calg_+^{eu} h_u(x,\mu)   {\rm Re}\big[C_{lequ}^{(3)}e^{-i\phi_-}\big] \Bigg).
\end{align}
The contributions from the anti-quark can be obtained from Eqs.\,\eqref{DSA:SMEFT1} and \eqref{DSA:SMEFT2} by the following replacement,
\begin{align}
	\label{eq:sub:C}
	C_{ledq} \rightarrow -C_{ledq},\quad C_{lequ}^{(1)}\rightarrow -C_{lequ}^{(1)},\quad \calg_{+}^{eq}\leftrightarrow \calg_{-}^{eq}. 
\end{align}
However, their contributions are negligible due to the suppression of the anti-quark transversity distributions~\cite{Zeng:2023nnb,Gamberg:2022kdb}. We also observe that the transverse DSA from the operator $\calO_{l e d q}$, which exhibits opposite chiral projections between lepton and quark bilinears, is sensitive to $\phi_+$, while the contributions from $\calO_{l e q u}^{(1,3)}$, which possess the same chiral structures of lepton and quark currents, would be sensitive to $\phi_-$. This behavior can be understood from the parity transformation of the lepton or quark current of operators.
Furthermore, the distinct behaviors of $\calO_{l e q u}^{(1,3)}$ on variable $y$ in Eqs.~\eqref{DSA:SMEFT1} and ~\eqref{DSA:SMEFT2} can be understood from the following Fierz identity of Dirac spinors\,\cite{Nieves:2003in,Liao:2012uj}:
\begin{align}
    \nonumber
    \label{eq:fierz}
    &~~(\overline{u}_{1}\sigma^{\mu\nu}P_{R} u_{2})(\overline{u}_{3}\sigma_{\mu\nu}P_{R} u_{4})=
    \\
    &-4(\overline{u}_{1}P_{R} u_{2})(\overline{u}_{3}P_{R} u_{4})+8(\overline{u}_{1}P_{R} u_{4})(\overline{u}_{3}P_{R} u_{2}).
\end{align}
Therefore, the contribution of $\calO_{l e q u}^{(3)}$ to the transverse DSA can be equivalently described by two scalar type operators according to Eq.\,\eqref{eq:fierz}.  The first interaction corresponds to the operator $\calO_{l e q u}^{(1)}$, which is proportional to $-4y$ (see Eqs.~\eqref{DSA:SMEFT1} and~\eqref{DSA:SMEFT2}), while the new scalar interaction will introduce a new structure $8(y-1)$. Thus, the total contribution from $\calO_{l e q u}^{(3)}$ will result in a large constant contribution compared to $\calO_{l e q u}^{(1)}$, causing a sign flip and significant enhancement in the absolute value of the DSA, given that $\abs{y}\leq 1$.

\begin{figure}[t!]
	\hspace{-0.5cm}
	\includegraphics[width=0.5\textwidth]{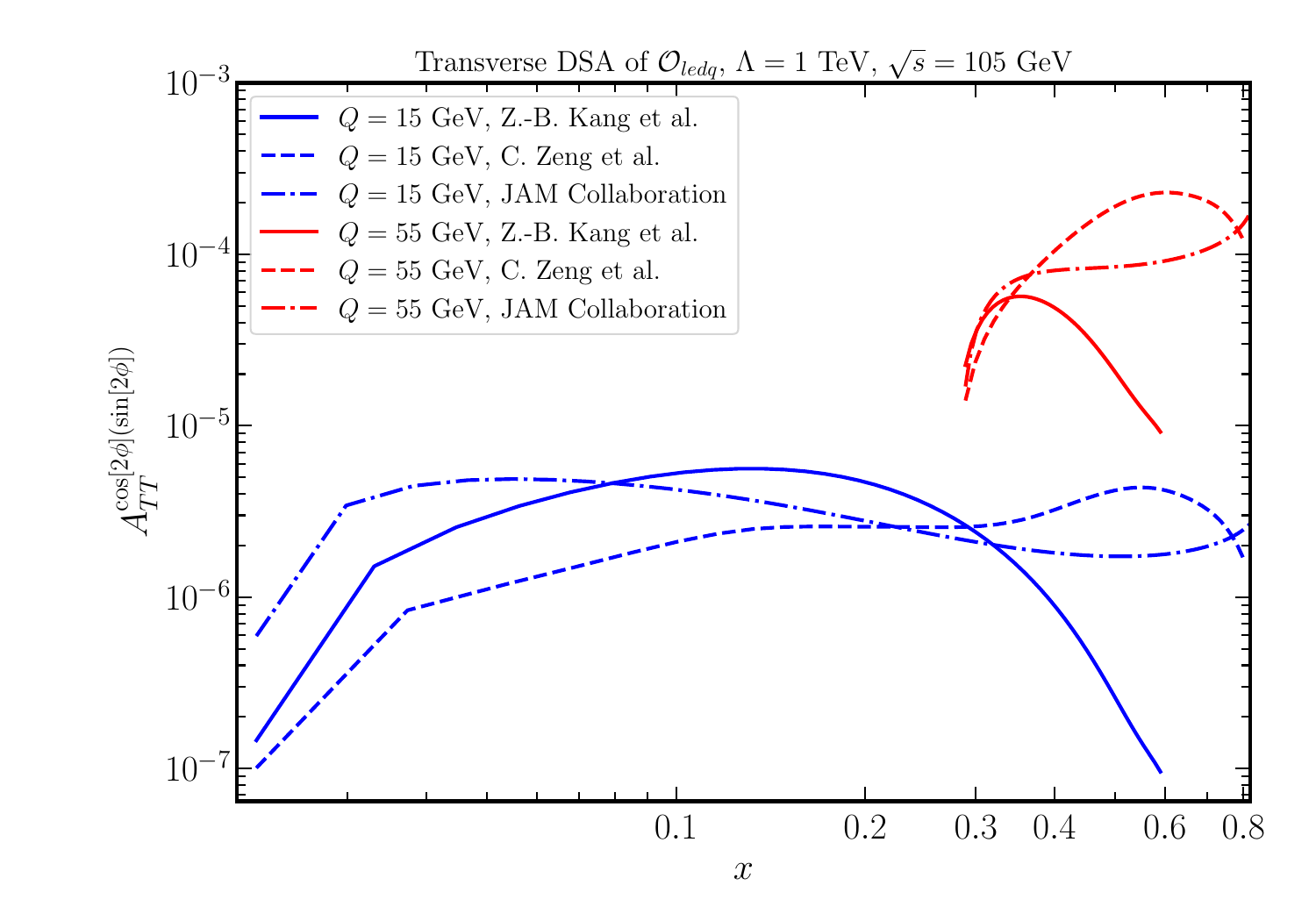}
	\caption{The integrated asymmetry $A_{TT}^w$ for operator $\calO_{leqd}$ with ${\rm Re}({\rm Im})[C_{ledq}]/\Lambda^2=1/{\rm TeV}^2$ at $\sqrt{S}=105$ GeV EIC. The solid lines depict the results obtained using the transversities from Ref.\,\cite{Kang:2015msa}. The dashed lines correspond to the transversities from Ref.\,\cite{Zeng:2023nnb}, while the dot-dashed lines represent the transersities from Ref.\,\cite{Gamberg:2022kdb}.}
	\label{fig:O1}
\end{figure}

\begin{figure*}[htbp]
\begin{centering}
\includegraphics[width=0.49 \textwidth]{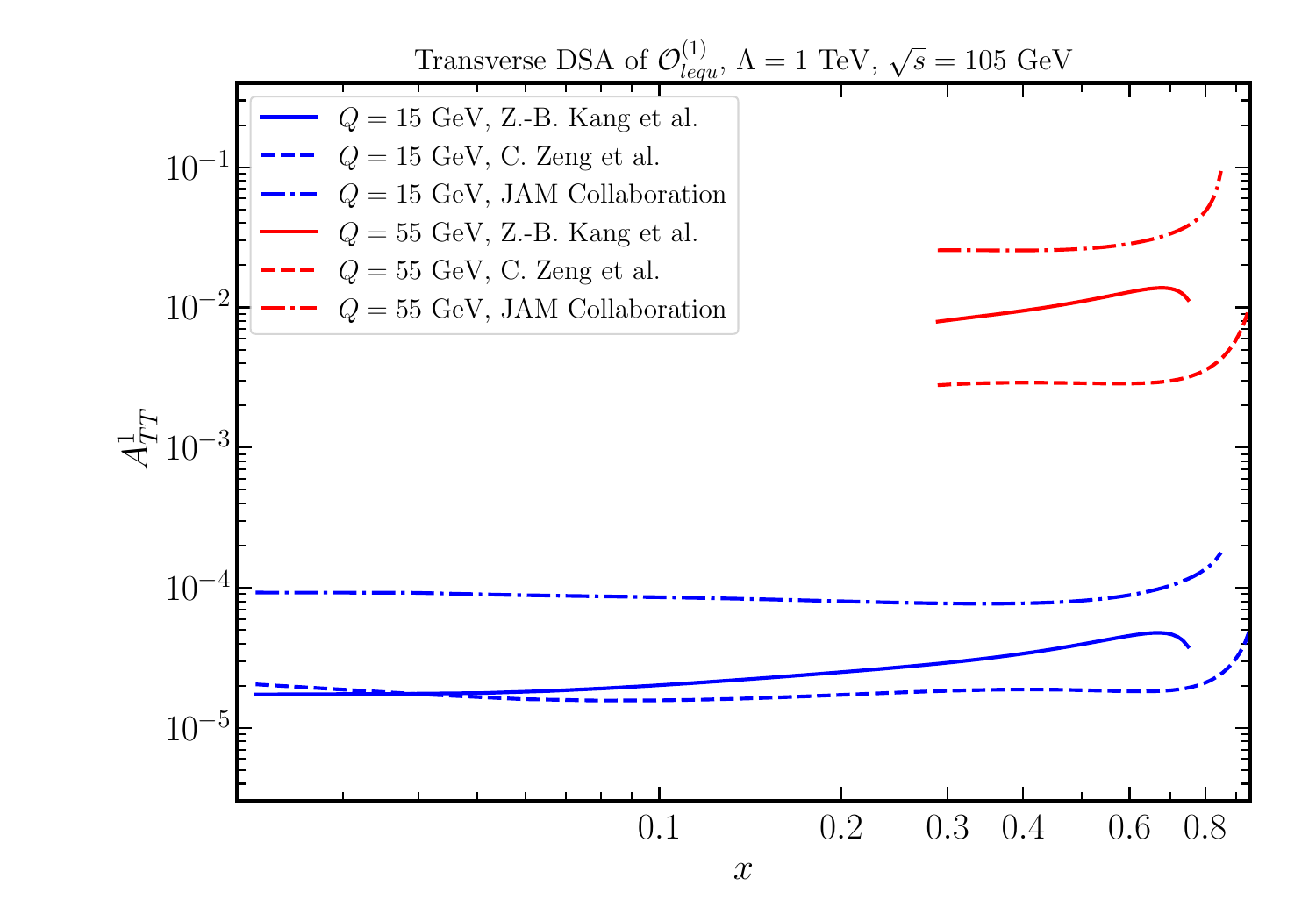}
\includegraphics[width=0.49 \textwidth]{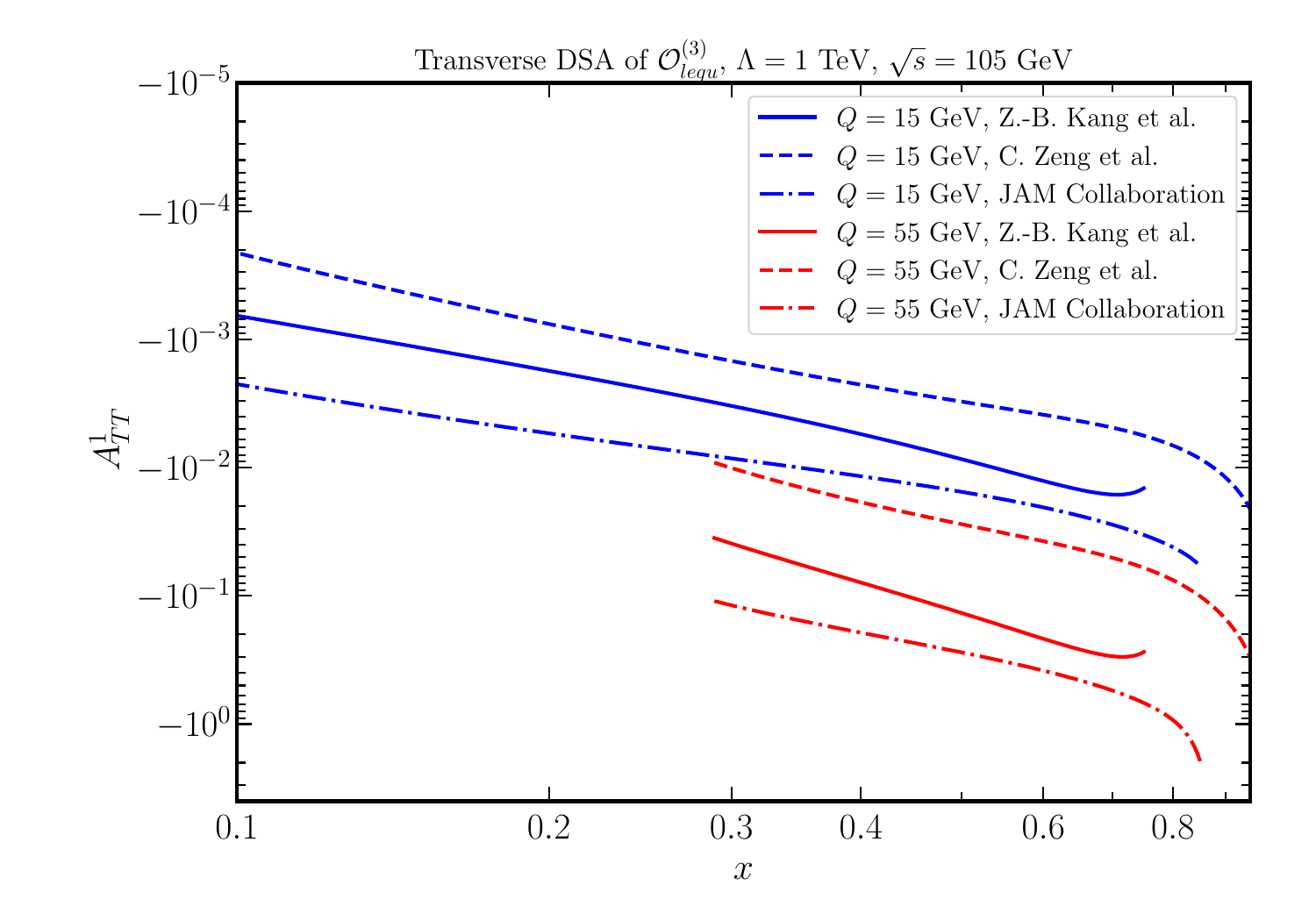}
\caption{Similar to Fig.~\ref{fig:O1}, but for operators $\calO_{lequ}^{(1,3)}$ with a weight function $w(\phi)=1$.
}
\label{fig:O2O3}
\end{centering}
\end{figure*}

Given the simple azimuthal dependence from the NP in Eqs.~\eqref{DSA:SMEFT1} and ~\eqref{DSA:SMEFT2}, it is sufficient to consider two experimental setups, (1) aligned or opposite spin setup: $\phi_1=\phi_2=\phi$ or $\phi_1=\phi_2+\pi=\phi$, and (2) perpendicular spin setup: $\phi_1=\phi_2+\pi/2=\phi$. Consequently, the transverse DSA from the operator $\calO_{l e d q}$ will exhibit a $\cos2\phi$ and $\sin2\phi$ behavior for its real and imaginary parts, respectively, but only flat distribution survives for the real part of $\calO_{l e q u}^{(1,3)}$ for the aligned/opposite spin setup. The behaviors for the operator $\calO_{l e d q}$ and $\calO_{l e q u}^{(1,3)}$ will be switched between their real and imaginary parts when considering the perpendicular spin configuration, i.e., the behavior will be switched for the real and imaginary parts of $C_{l e d q}$, while only a flat distribution survives for imaginary part of $C_{l e q u}^{(1,3)}$.

To extract the information about the scalar/tensor four-fermion operators, we can define the weight-integrated asymmetry as,
\begin{align}
	A_{TT}^w=\frac{1}{2\pi}\int_0^{2 \pi} d \phi~w(\phi) A_{TT}(\phi),
 \label{eq:ATTW}
\end{align}
where $\omega(\phi)$ is the weight function to project out the different parts of these NP effects via the azimuthal dependence of the $A_{TT}$. Based on Eqs.\,\eqref{DSA:SMEFT1} and \eqref{DSA:SMEFT2}, it is clear that the contributions of $\re[C_{ledq}]$ and $\im[C_{ledq}]$ can be isolated by the weight function $w(\phi)=\cos2\phi$ and $w(\phi)=\sin2\phi$ respectively, while we can choose $w(\phi)=1$ to simultaneously capture the contributions of ${\rm Re}\big[C_{lequ}^{(1)}\big]$ and ${\rm Re}\big[C_{lequ}^{(3)}\big]$ for the aligned/opposite spin setup.  The imaginary effects of $\calO_{l e q u}^{(1,3)}$ can be obtained when considering the perpendicular spin configuration.

In Figs.\,\ref{fig:O1} and \,\ref{fig:O2O3}, we present the integrated asymmetries arising from the three chirality-flipping four-fermion operators with the Wilson coefficients $C_i=1$ and the NP scale $\Lambda=1$ TeV for three different transversity distributions in Refs.~\cite{Kang:2015msa,Zeng:2023nnb,Gamberg:2022kdb} at the $\sqrt{S}=105$ GeV EIC with aligned spin setup ($\phi_1=\phi_2=\phi$). As expected, the transverse DSAs from these NP operators are significantly enhanced compared to the prediction in the SM, particularly for the tensor type operator $\calO_{l e q u}^{(3)}$ which could generate an asymmetry on the order of magnitude of $\mathcal{O}(10^{-3})\sim\mathcal{O}(1)$. It makes this method promising for probing the scalar/tensor four-fermion operators at the EIC/EicC. Additionally, we observe that the integrated asymmetries from $\calO_{ledq}$ and $\calO_{l e q u}^{(1)}$ are positive, while a negative asymmetry is generated by $\calO_{l e q u}^{(3)}$, and its absolute value is much larger than the predictions from $\calO_{ledq}$ and $\calO_{l e q u}^{(1)}$, and it is much more sensitive to the Bjorken-$x$. This behavior arising from the large contribution of the constant factor from the tensor operator, as we discussed before, see Eqs.~\eqref{DSA:SMEFT1} and~\eqref{DSA:SMEFT2}. Moreover, the scalar/tensor four-fermion operators increase significantly with respect to the transfer energy $Q$, which is the result of the $\calO(Q^2 / \Lambda^2)$ behavior of these operators to the observables.

\section{Sensitivity at EIC and EicC}
\label{sec:sensitivity}

\begin{table}
	\centering
	\renewcommand\arraystretch{1.2}
	\begin{tabular}{|c|c|c|} 
        \hline 
		\multirow{2}{*}{Transversity} & \multicolumn{2}{|c|}{ Limits on $\Re[C_{ledq}] (\Im[C_{ledq}])$} \\
        \cline{2-3}
		   &  EIC ($105$ GeV) & EicC ($16.7$ GeV) \\
		\hline
		 Z.-B. Kang et al~\cite{Kang:2015msa} & 5.16 & 34.60 \\
		\hline
        C. Zeng et al~\cite{Zeng:2023nnb} & 4.53 & 13.72 \\
		\hline
        JAM Collaboration~\cite{Gamberg:2022kdb} & 5.12 & 29.69 \\
		\hline
	\end{tabular}
	\caption{The projected sensitivities of probing operator $\calO_{ledq}$ from the DSA measurement at EIC and EicC at 68\% C.L. with the integrated luminosity $\mathcal{L}=100~{\rm fb}^{-1}$, assuming $\Lambda=1~{\rm TeV}$.}
	\label{tab:con}
\end{table}

\begin{figure*}[htbp]
\begin{centering}
\includegraphics[width=0.45 \textwidth]{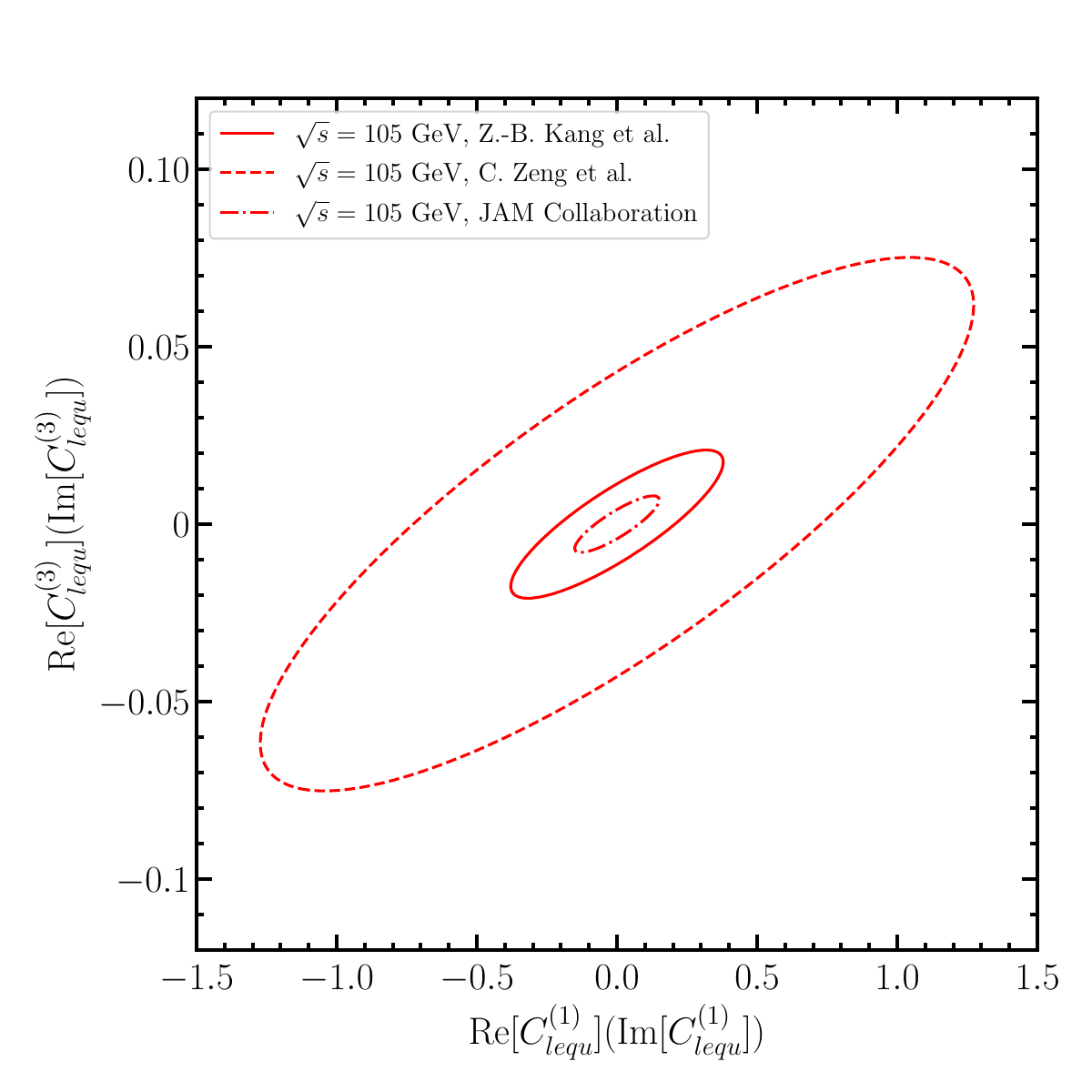}
\includegraphics[width=0.45 \textwidth]{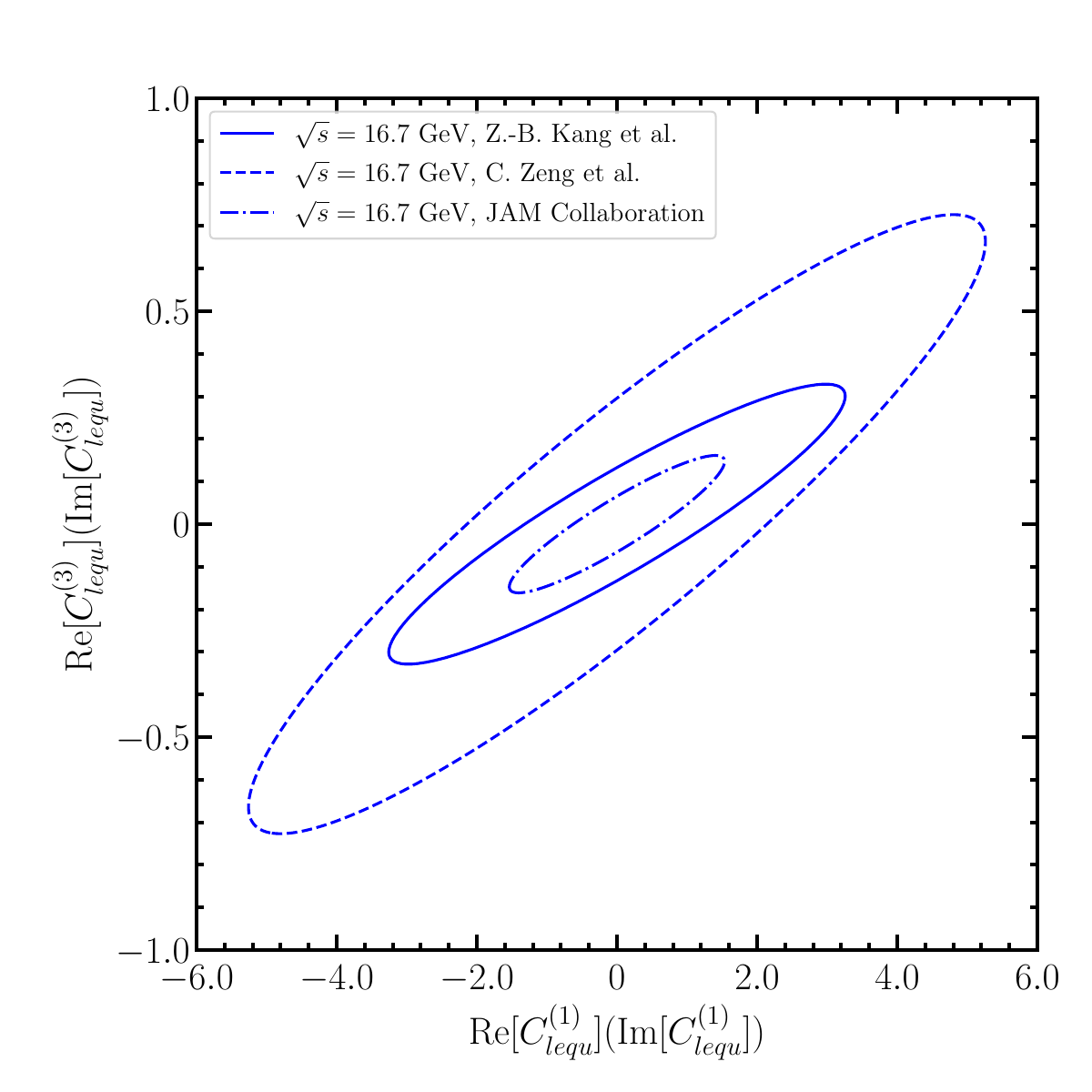}
\caption{The expected limits on the Wilson coefficients ${\rm Re}[C_{lequ}^{(1)}]$ (${\rm Im}[C_{lequ}^{(1)}]$) and ${\rm Re}[C_{lequ}^{(3)}]$ (${\rm Im}[C_{lequ}^{(3)}]$) from the DSA measurements at the EIC (left) and EicC (right) at 68\% C.L. with the aligned/opposite (perpendicular) spin configuration of electron and proton, assuming $\mathcal{L}=100~{\rm fb}^{-1}$ and $\Lambda=1~{\rm TeV}$. The solid, dashed, and dot-dashed contours represent the limits obtained with the transversities in Refs. \cite{Kang:2015msa}, \cite{Zeng:2023nnb}, and \cite{Gamberg:2022kdb} respectively. }
\label{fig:combine-lim}
\end{centering}
\end{figure*}

In this section, we estimate the expected sensitivities of probing the scalar/tensor four-fermion operators through the transverse DSAs at the EIC/EicC under the aligned spin ($\phi_1=\phi_2=\phi$) and perpendicular spin configurations ($\phi_1=\phi_2+\pi/2=\phi$). The weight-integrated asymmetry in Eq.~\eqref{eq:ATTW} can be translated to the experimental measurements,
 \begin{align}
 \label{eq:asy_observale}
 	\nonumber
 	&A_{TT}^w=\frac{1}{P_{T,e}P_{T,p}}\frac{1}{N_{\uparrow\uparrow}+N_{\downarrow\downarrow}+N_{\uparrow\downarrow}+N_{\downarrow\uparrow}}
 	\\
     \nonumber
 	&\times\int_{0}^{2\pi} d\phi w(\phi)\Big( N_{\uparrow\uparrow}(\phi)+N_{\downarrow\downarrow}(\phi)
     \\
     &\qquad\qquad\qquad\qquad\quad-N_{\uparrow\downarrow}(\phi)-N_{\downarrow\uparrow}(\phi)\Big),
 \end{align}
where $N_{ij}$ denotes the measured event number with specific transversely polarized configuration of electron and proton. 

It has been demonstrated that the maximization of the integrated luminosity has a more substantial impact on the sensitivity of probing the SMEFT effects than a slight increase in collider energy at the EIC~\cite{Boughezal:2023ooo,Boughezal:2022pmb}. As a result, we will consider $\sqrt{S}=105$ GeV at the EIC as a benchmark collider energy for probing the NP effects, since it is expected to achieve the highest luminosity~\cite{AbdulKhalek:2021gbh}.  In our analysis, we focus on the following kinematic region: $x\in[0.1,0.8]$ and $Q\in[15, 65]$ GeV, incorporating an inelasticity cut of $0.01 \leq y \leq 0.95$ \cite{AbdulKhalek:2021gbh}. A similar analysis will also be done for the EicC with a smaller collider energy $\sqrt{S}= 16.7$ GeV and $Q\in[6, 11]$ GeV, applying the same cuts on $x$ and $y$ variables as for the EIC~\cite{Anderle:2021wcy}. 
We consider the integrated asymmetry $A_{TT}^\omega$ in separate bins in $(Q,x)$ space, assuming that the statistical errors, given by Eq.~\eqref{eq:asy_unc}, are comparable to the values of $A_{TT}^\omega$ and slowly change as the variable $x$. Consequently, we can ignore the systematic uncertainties for the asymmetries~\cite{ZEUS:2009swh,Yan:2022npz}, allowing us to use the theoretical predictions from Fig.~\ref{fig:O1} and \ref{fig:O2O3} directly. We then conduct a $\chi^2$ analysis to constrain the Wilson couplings,
\begin{align}
	\chi^2=\sum_i\left[\frac{A_i^{\mathrm{th}}-A_i^{\mathrm{exp}}}{\delta A_i}\right]^2,
\end{align}
where $A_i^{\mathrm{th}}$ and $A_i^{\mathrm{exp}}$ represent, respectively, the theoretical prediction of asymmetry
induced by the four-fermion operators and the experimentally measured value for the $i$-th $(Q,x)$-bin following Eq.~\eqref{eq:asy_observale}. 
For simplicity, we have assumed the experimental values are consistent with the SM predictions, and are negligible. $\delta A_i$ denotes the corresponding statistical uncertainties of the $i$-th bin. The statistical uncertainty for $A_{TT}^w$ is given by
\begin{align}
\label{eq:asy_unc}
	\delta A_{TT}^w&\simeq \frac{1/(P_{T,e} P_{T,p})}{\sqrt{4{\cal L}\sigma(P_{T,e(p)}=0)}}\cdot \sqrt{\frac{\int_{0}^{2\pi} d\phi w^2(\phi)}{2\pi}},
\end{align}
where ${\cal L}$ is the integrated luminosity of each dataset with a specific transversely polarized configuration, assumed to be the same for each one.
We should note that the number of bins is not fixed for all analyses, as it depends on the collider energy, transversities, and operators.

In Table~\ref{tab:con}, we present the expected constraining power of EIC/EicC on the scalar operator $\mathcal{O}_{l e d q}$, assuming $\Lambda=1~{\rm TeV}$ and $P_{T,e}=P_{T,p}=0.7$, with the canonical integrated luminosity, $\mathcal{L}=100~{\rm fb}^{-1}$ at $68\%$ confidence level (C.L.). The weight function $w(\phi)=\cos2\phi$ and $\sin2\phi$ have been used to constrain its real and imaginary parts of the Wilson coefficient. A similar constraint for operators $\mathcal{O}_{l e q u}^{(1,3)}$ can be obtained by choosing $w(\phi)=1$, and the results are shown in Fig.~\ref{fig:combine-lim}. It shows that we can constrain their real and imaginary parts under the aligned and perpendicular spin configurations, and the parameter space for the real and imaginary coefficients is exactly the same.  We also find that the ability for EIC to constrain the chirality-flipping four-fermion operators is much better than EicC due to the much higher collider energy. The sensitivity to the operators involving $u$-quark is much better than $d$-quark, because of the larger electric charge and non-perturbative quark transversity.

However, it is important to note that these conclusions depend strongly on the quark transversity distributions, which are currently poorly constrained but could be determined with high accuracy in the upcoming EIC/EicC. As a result, our findings still offer promising avenues for probing chirality-flipping semi-leptonic four-fermion operators, particularly for the operator involving the $u$-quark. Importantly, the conclusions would not be sensitive to other potential NP effects in the DIS process, such as the vector and axial-vector type four-fermion operators.

We also note that these operators can contribute to the cross-section of Drell-Yan process at the LHC and low-energy experiments at $\mathcal{O}(1/\Lambda^4)$ with a specific flavor assumption~\cite{Boughezal:2021tih}. It was found that $C_i\sim \mathcal{O}(0.01)$ from Drell-Yan data when one operator is considered at a time~\cite{Boughezal:2021tih}. The limits from low-energy measurements, such as nuclear beta decays, the ratio between $\Gamma(\pi^+\to e^+\nu)$ and $\Gamma(\pi^+\to \mu^+\nu)$, and radiative pion decays, are comparable to the LHC measurement. However, the cross section alone is challenging to disentangle the scalar and tensor operators from other NP effects, especially the dim-8 operators, which will also contribute to the cross section at $\calO(1/\Lambda^4)$, and could complicate the analysis and weaken the conclusions~\cite{Alioli:2020kez}. Additionally, it also lacks the sensitivity to distinguish the real and imaginary parts of these operators through the traditional methods, making the DSAs at the transversely polarized EIC/EicC a unique opportunity to probe these NP effects, which is complementary and competitive with respect to other methods in the literature.

\section{Conclusion}
\label{sec:conclusion}
In this paper, we propose using the transverse double spin asymmetry observables to investigate the semi-leptonic scalar/tensor type four-fermion operators involving electrons and light quarks at the future Electron-Ion Collider with transversely polarized electron and proton beams. Due to the double fermion helicity flip associated with these operators, we demonstrated that the interference of these NP effects with the SM will generate sizeable DSAs and could produce distinct $\cos2\phi$ and $\sin2\phi$ distributions. Notably, this occurs without the suppression of the electron and light quark masses at $\calO(1/\Lambda^2)$ and without the contamination from the SM and other potential NP effects. Consequently, the anticipated limits for these operators using this method are expected to be stronger or comparable to those obtained through other approaches in Drell-Yan processes at the LHC, which can only occur at $\calO(1/\Lambda^4)$ in the massless limit of fermions. This is particularly true for the tensor type four-fermion operator of $u$ quark. Importantly, our approach provides the opportunity to simultaneously constrain the real and imaginary parts of those couplings, enabling direct study of potential CP-violating effects arising from these operators. Additionally, we found that these results are strongly dependent on the transversity distribution of quarks, which is currently poorly constrained by experimental data. However, the knowledge of these non-perturbative functions is expected to be significantly improved in the upcoming Electron-Ion Collider. Thus, our approach is expected to play a crucial role in probing these NP effects in the future.

\section*{acknowledgments}
H. L. Wang and H. Xing are supported by the Guangdong Major Project of Basic and Applied Basic Research No. 2020B0301030008 and No. 2022A1515010683, by the National Natural Science Foundation of China under Grants No. 12247151, No. 12022512 and No. 12035007. X.-K. Wen is supported in part by the National Science Foundation of China under Grants No.11725520, No.11675002 and No.12235001.
B. Yan is supported by the IHEP under Contract No. E25153U1.  The authors thank Z. B. Kang and Y. Y. Zhou for sharing the transversity distributions in Ref.~\cite{Kang:2015msa} and H. X. Dong and P. Sun for sharing the transversely polarized PDFs in Ref.~\cite{Zeng:2023nnb}. We thank D. Pitonyak for bringing the most updated global analysis of transversity PDF into our attention~\cite{Gamberg:2022kdb,Cocuzza:2023oam,Cocuzza:2023vqs}.

\bibliographystyle{apsrev}
\bibliography{references-paper}

\end{document}